\documentclass[11pt]{article}
\newcommand{\bq}{\begin{equation}}
\newcommand{\eq}{\end{equation}}
\newcommand{\ba}{\begin{eqnarray}}
\newcommand{\ea}{\end{eqnarray}}

\newcommand{\nl }{ \nonumber  }

\newcommand{\p}{\partial}
\newcommand{\h}{\hspace{.5cm}}
\newcommand{\s}{\sigma}

\newcommand{\la}{\lambda}
\newcommand{\La}{\Lambda}
\newcommand{\Di}{\left(\p_0-\la^{i}\p_i\right)}
\newcommand{\Dj}{\left(\p_0-\la^{j}\p_j\right)}

\def\appendix#1{
  \setcounter{equation}{0}
  \renewcommand{\thesection}{\Alph{section}}
  \section*{Appendix \thesection\protect\indent \parbox[t]{11.15cm}
  {#1} }
  \addcontentsline{toc}{section}{Appendix \thesection\ \ \ #1}
  }

\begin{document}
\vspace*{2cm}
\begin{center}
{\bf $M2$-BRANE SOLUTIONS IN $AdS_7\times S^4$
\vspace*{0.5cm}\\ P. Bozhilov}
\\ {\it Department of Theoretical and Applied Physics, \\
Shoumen University, 9712 Shoumen, Bulgaria\\
E-mail:} p.bozhilov@shu-bg.net\\
and\\
{\it The Abdus Salam International Centre for Theoretical Physics,
Trieste, Italy\\
E-mail:} pbozhilo@ictp.trieste.it\\
\end{center}
\vspace*{0.5cm}

We consider different M2-brane configurations in the M-theory
$AdS_7\times S^4$ background, with field theory dual $A_{N-1}(2,0)$ $SCFT$.
New membrane solutions are found and compared
with the recently obtained ones.

\vspace*{.5cm}
\vspace*{.5cm}

\section{Introduction}

The paper \cite{4} by Gubser, Klebanov and Polyakov on the semi-classical
limit of the $AdS/CFT$ duality has inspired a lot of interest in the
investigation of the existing connections between the classical string
solutions, their semi-classically quantized versions and the relevant objects
on the field theory side. Different string configurations have been considered,
describing rotating, pulsating or orbiting strings. Much attention has been
paid to string solutions in type IIB $AdS_5\times S^5$ background with field
theory dual $\mathcal{N}=4$ $SU(N)$ $SYM$ in four dimensional flat space-time.
Moreover, the string dynamics has been investigated in many other string theory
backgrounds, known to have field theory dual descriptions in
different dimensions, with different number of (or without)
supersymmetries, conformal or non-conformal. Besides, membrane solutions in
M-theory backgrounds have been obtained \cite{6} - \cite{27}.
In \cite{6} - \cite{14}, M2-brane configurations have been considered in
$AdS_7\times S^4$ space-time, with field theory dual $A_{N-1}(2,0)$ $SCFT$.
Rotating membrane solution in $AdS_7$ have been obtained in \cite{6}.
Rotating and boosted membrane configuration was investigated in \cite{10}.
Multiwrapped circular membrane, pulsating in the radial direction of $AdS_7$,
has been considered in \cite{14}. The article \cite{27} is devoted to the
investigation of rotating membranes on $G_2$ manifolds.

Here, we will be interested in obtaining new membrane solutions in
$AdS_7\times S^4$ M-theory background. In section 2, we give brief
description of the recently received M2-brane solutions in this
space-time. In section 3, we firstly settle the framework, which
we will work in.\footnote{Actually, we will use the general
approach developed in \cite{NPB656}.} Then, we proceed to find
several M2-brane solutions, based on two different types of
membrane embedding. The generic formulas, necessary for our
calculations in this section, are collected in appendix.

\section{Short review of the recent M2-brane solutions in $AdS_7\times S^4$}

Let us review briefly some of the results obtained recently in
\cite{6} - \cite{14}, concerning the M2-brane dynamics on $AdS_7\times S^4$
background.

In \cite{6}, a rotating membrane in $AdS_7$ was considered. The background
metric is taken in global coordinates
\ba\nl ds^2_{AdS_{7}} = -\cosh^2\rho dt^2 + d\rho^2 + \sinh^2\rho
\left[d\theta^2 + \sin^2\theta\left(d\phi^2+\sin^2\phi
d\Omega^2_3\right)\right].\ea The M2-brane worldvolume coordinates are
$(\tau,\s,\varphi)$ and the rotating membrane configuration is given by the
ansatz
\ba\label{a6} t=\tau,\h\rho=\rho(\s),\h\theta=\theta(\varphi),
\h\phi=\omega\tau.\ea
Therefore, the metric seen by the M2-brane is
\ba\nl ds^2 = -\cosh^2\rho dt^2 + d\rho^2 + \sinh^2\rho
\left(d\theta^2 + \sin^2\theta d\phi^2\right) .\ea
This background does not depend on the coordinates $t$ and $\phi$, which leads
to the conservation of the corresponding generalized momenta - the energy and
the spin of the membrane. For the configuration (\ref{a6}), they were found
to be
\ba\nl &&E=4N\int_0^{\rho_0}d\rho\int_{\theta_1}^{\theta_2}d\theta
\frac{\cosh^2\rho\sinh\rho}{\sqrt{\cosh^2\rho
- \omega^2\sinh^2\rho\sin^2\theta}},\\ \nl
&&s= 4N\omega\int_0^{\rho_0}d\rho\int_{\theta_1}^{\theta_2}d\theta
\frac{\sinh^3\rho\sin^2\theta}{\sqrt{\cosh^2\rho
- \omega^2\sinh^2\rho\sin^2\theta}}.\ea

Rotating and boosted membrane configuration was investigated in \cite{10}.
The following coordinates for the $AdS_7\times S^4$ metric have been used
\ba\nl l_p^{-2}ds^2_{AdS_{7}\times S^4} &=& 4R^2\left\{-\cosh^2\rho dt^2 +
d\rho^2 + \sinh^2\rho\left(d\psi_1^2 + \cos^2\psi_1 d\psi_2^2+\sin^2\psi_1
d\Omega^2_3\right)\right.\\ \label{AdS7S4m} &+&\left.\frac{1}{4}\left[d\alpha^2
+ \cos^2\alpha d\theta^2+\sin^2\alpha\left(d\beta^2 + \cos^2\beta d\gamma^2
\right)\right]\right\},\\ \nl
d\Omega^2_3 &=& d\psi_3^2 + \cos^2\psi_3 d\psi_4^2 +
\cos^2\psi_3\cos^2\psi_4 d\psi_5^2,\h R^3=\pi N.\ea
The coordinates which parameterize the membrane worldvolume are chosen to be
$(\xi_1,\xi_2,\xi_3) = (\tau,\delta,\s)$. Then, the considered M2-brane
embedding can be written as follows
\ba\label{a10} t=\kappa\tau,\h \rho=\rho(\s),\h \psi_1=\pi/4,
\h \psi_2=\sqrt{2}a\delta,
\h\psi_5=\sqrt{2}\omega\tau,\h \theta=2\nu\tau,\ea
and all other coordinates set to zero. Hence, the background felt by the
membrane is
\ba\nl ds^2 = (2l_p R)^2\left[-\cosh^2\rho dt^2 +
d\rho^2 + \frac{1}{2}\sinh^2\rho\left(d\psi_2^2+d\psi_5^2\right)
+\frac{1}{4}d\theta^2\right].\ea
This metric does not depend on four coordinates - $t$, $\psi_2$, $\psi_5$ and
$\theta$. The conserved quantities, corresponding to the Killing vectors
$\p/\p t$, $\p/\p\psi_5$ and $\p/\p\theta$, have been obtained
to be given by the equalities
\ba\nl &&E= \frac{4R^3}{\pi}\kappa\int_0^{\rho_0}
\frac{\sinh\rho\cosh^2\rho d\rho}{\sqrt{(\kappa^2-\nu^2)\cosh^2\rho
- (\omega^2-\nu^2)\sinh^2\rho}},\\ \nl
&&S= \frac{4R^3}{\pi}\omega\int_0^{\rho_0}
\frac{\sinh^3\rho d\rho}{\sqrt{(\kappa^2-\nu^2)\cosh^2\rho
- (\omega^2-\nu^2)\sinh^2\rho}},\\ \nl
&&J= \frac{4R^3}{\pi}\nu\int_0^{\rho_0}
\frac{\sinh\rho d\rho}{\sqrt{(\kappa^2-\nu^2)\cosh^2\rho
- (\omega^2-\nu^2)\sinh^2\rho}}.\ea
It was pointed out in \cite{10} that there exists the following connection
between the energy $E$, the spin $S$ and the $R$-charge $J$ of the membrane
\ba\nl E=\frac{\kappa}{\omega}S+\frac{\kappa}{\nu}J.\ea
Then, this constraint has been used to determine the dependence of $E$ on
$S$ and $J$.

Another type of M2-brane configuration - multiwrapped circular membrane
pulsating in the radial direction of $AdS_7$, has been considered in \cite{14}.
The coordinates on $AdS_7\times S^4$ and on the M2-brane worldvolume are
chosen as in \cite{10}. The membrane embedding is given by the ansatz
\ba\label{a14} t=\tau,\h \rho=\rho(\tau),\h \psi_1=\pi/4,\h
\psi_2=\sqrt{2}a\delta, \h\psi_5=\sqrt{2}m\s.\ea
It follows from here that the metric seen by the M2-brane is
\ba\label{rb14} ds^2 = (2l_p R)^2\left[-\cosh^2\rho dt^2 +
d\rho^2 + \frac{1}{2}\sinh^2\rho\left(d\psi_2^2+d\psi_5^2\right)\right].\ea
The relevant action for such membrane configuration reads \cite{14}:
\ba\label{ma14} I=-\left(2R\right)^3 am\int dt \sinh^2\rho
\sqrt{\cosh^2\rho - \dot{\rho}^2}.\ea

\setcounter{equation}{0}
\section{New M2-brane solutions in $AdS_7\times S^4$}

In considering the M2-brane dynamics, we will use the following
action for a membrane moving in curved space-time with metric tensor
$g_{MN}(x)$, and interacting with a background 3-form gauge field $b_{MNP}(x)$
\ba\label{oma} S &=&\int
d^{3}\xi\mathcal{L}= \int
d^{3}\xi\left\{\frac{1}{4\lambda^0}\Bigl[g_{MN}\left(X\right)
\left(\p_0-\lambda^i\p_i\right) X^M\left(\p_0-\lambda^j\p_j\right)X^N \right.
\\ \nl &-& \left.
\left(2\lambda^0T_2\right)^2\det{\left(g_{MN}(X)\p_i X^M\p_j X^N\right)}
\Bigr] + T_2 b_{MNP}(X)\p_0X^{M}\p_1X^N\p_2X^{P} \right\},
\\ \nl && \xi = (\xi^0,\xi^1,\xi^2)=(\tau,\delta,\s),\h
\p_m=\p/\p\xi^m,\\ \nl &&m = (0,i) = (0,1,2),\h M = (0,1,\ldots,10),\ea
where $\lambda^m$ are Lagrange multipliers, $x^M=X^M(\xi)$ are the membrane
embedding coordinates, and $T_2$ is its tension. This action is
classically equivalent to the Nambu-Goto type action \footnote{
Namely this action has been used in \cite{6} - \cite{14}.}
\ba\label{ngma} S^{NG}= - T_2\int d^{3}\xi\left[
\sqrt{-\det{\left(\p_m X^M \p_n X^N g_{MN}(X)\right)}}
-\frac{1}{6}\varepsilon^{mnp}
\p_{m}X^{M}\p_n X^N \p_{p}X^{P} b_{MNP}(X)\right],\ea
and to the Polyakov type action
\ba\nl S^{P}= - \frac{T_2}{2}\int
d^{3}\xi\left\{\sqrt{-\gamma}\left[\gamma^{mn} \p_m X^M\p_n
X^N g_{MN}(X)-1\right]\right. \\ \nl - \left.\frac{1}{3} \varepsilon^{mnp}
\p_{m}X^{M}\p_nX^N\p_{p}X^{P}b_{MNP}(X)\right\},\ea
as shown in \cite{NPB656}.

We choose to work with the action (\ref{oma}), because it possesses the
following advantages. First of all, it does not contain square root,
thus avoiding the introduction of additional nonlinearities in the equations
of motion. Besides, the equations of motion for the Lagrange multipliers
$\lambda^m$ generate the {\it independent} constraints only
\ba\label{con0} &&G_{00}-2\lambda^{j}G_{0j}+\lambda^{i}\lambda^{j}G_{ij}
+\left(2\lambda^0T_2\right)^2\det{\left(G_{ij}\right)}=0,\\
\nl &&G_{0j}-\lambda^{i}G_{ij}=0,\ea where
\ba\label{im} G_{mn}=\p_m X^M\p_n X^N g_{MN}(X)\ea is the
metric induced on the membrane worldvolume. Finally, this action gives a
unified description for the tensile and tensionless membranes, so the limit
$T_2\to 0$ may be taken at any stage of our considerations.

Further on, we will use the gauge $\lambda^m=constants$, in which the
equations of motion for $X^M$, following from (\ref{oma}), are given by
$(\mathbf{G}\equiv\det{\left(G_{ij}\right)})$
\ba\nl &&g_{LN}\left[\Di\Dj X^N - \left(2\lambda^0T_2\right)^2
\p_i\left(\mathbf{G}G^{ij}\p_j X^N\right)\right]\\ \nl
&&+\Gamma_{L,MN}\left[\Di X^M \Dj X^N - \left(2\lambda^0T_2\right)^2
\mathbf{G}G^{ij}\p_i X^M \p_j X^N\right]\\ \nl
&&=2\la^0 T_2 H_{LMNP}\p_0X^{M}\p_1X^N\p_2X^{P},\ea
where
\ba\nl \Gamma_{L,MN}=g_{LK}\Gamma^K_{MN}=\frac{1}{2}\left(\p_Mg_{NL}
+\p_Ng_{ML}-\p_Lg_{MN}\right)\ea
are the components of the symmetric connection compatible with the metric
$g_{MN}$ and $H_{LMNP}$ is the field strength of the
$3$-form gauge potential $b_{MNP}$.

We will investigate the M2-brane dynamics in the framework of
the following two types of embedding $(\Lambda^\mu_m = constants)$
\ba\label{tLA} X^\mu(\tau,\xi^i)=\Lambda^\mu_m \xi^m=
\Lambda^\mu_0\tau+\Lambda^\mu_1\delta+ \Lambda^\mu_2\s,\h
X^a(\tau,\xi^i)=Y^a(\tau),\ea and
\ba\label{tGA} X^\mu(\tau,\xi^i)=\Lambda^\mu_1\delta
+ \Lambda^\mu_2\s + Y^\mu(\tau),\h X^a(\tau,\xi^i)=Y^a(\tau).\ea
Here, the embedding coordinates $X^M(\tau,\xi^i)$ are divided into
$X^M=(X^\mu,X^a)$, where $X^\mu(\tau,\xi^i)$ correspond to the space-time
coordinates $x^\mu$, on which the background fields do not depend
\ba\label{ob} \p_\mu g_{MN} =0,\h \p_\mu b_{MNP} =0.\ea
In other words, we suppose that there exist $n_\mu$ commuting Killing vectors
$\p/\p x^\mu$, where $n_\mu$ is the number of the coordinates $x^\mu$.
The two ansatzes - (\ref{tLA}) and (\ref{tGA}), will be referred to as
{\it linear gauges} and {\it general gauges}, in analogy with the name
{\it static gauge} used for the embedding $X^m(\xi^n)=\xi^m$.

All formulas, necessary for our calculations in this section, are
given in appendix.

\subsection{Exact membrane solutions in linear gauges}

Comparing the M2-brane embeddings (\ref{tLA}) and (\ref{tGA}), which we are
going to explore, with the previously used ones (\ref{a6}), (\ref{a10}) and
(\ref{a14}), one sees that only (\ref{a14}) is of the same type. Namely, it is
particular case of (\ref{tLA}), corresponding to $(X^\mu=X^{0,2,3}, X^a=X^1)$
\ba\nl &&\Lambda_0^0=1,\h \Lambda_i^0=0,\h \Lambda_1^2=\sqrt{2}a,\h
\Lambda_0^2=\Lambda_2^2=0,\h\Lambda_2^3=\sqrt{2}m,\h\Lambda_0^3=\Lambda_1^3=0,
\\ \nl &&X^1(\tau,\delta,\s)=Y^1(\tau)=\rho(\tau).\ea

As far as classical membrane solution has been not given in \cite{14}, we
begin with obtaining such solution, based on their ansatz (\ref{a14}).
Let us first write down the two actions - (\ref{oma}) and (\ref{ngma}), for the
case under consideration. To this end, we need to compute the induced metric
(\ref{im}).  It can be found by comparing (\ref{con0}) with (\ref{c0l}), for
example. Its nonzero components are
\ba\nl G_{00}= -(2l_pR)^2\left(\cosh^2\rho-\dot{\rho}^2\right),\h
G_{11}=(2l_pR)^2a^2\sinh^2\rho,\h G_{22}=(2l_pR)^2m^2\sinh^2\rho.\ea
Taking this into account, one receives
\ba\nl S^{NG}=-(2\pi)^2T_2(2l_pR)^3am\int dt
\sinh^2\rho\sqrt{\cosh^2\rho-\dot{\rho}^2},\ea
which reproduces the Nambu-Goto type action (\ref{ma14}), used in \cite{14},
for
\ba\nl T_2=\frac{1}{(2\pi)^2 l_p^3}.\ea
Our action for this case is given by (\ref{olga}), and it reads
\ba\label{oelg1} S^{LG}&=&\frac{(2\pi l_p R)^2}{\la^0}\int dt
\left\{\dot{\rho}^2+\left[\left(\la^1a\right)^2+\left(\la^2m\right)^2\right]
\sinh^2\rho\right. \\ \nl &-&\left.\cosh^2\rho
-\left(2\la^0T_2\right)^2(2l_pR)^2a^2m^2\sinh^4\rho\right\}.\ea

Since our membrane configuration is defined by (\ref{a14}), the relevant
background is (\ref{rb14}). It does not depend on $x^0=t$, $x^2=\psi_2$ and
$x^3=\psi_5$, i.e. we have three commuting Killing vectors $\p/\p t$,
$\p/\p\psi_2$ and $\p/\p\psi_5$. Correspondingly, the Lagrangian in
(\ref{oelg1}) does not depend on $x^0=t$, $x^2=\psi_2$ and $x^3=\psi_5$.
Therefore, the conjugated momenta $P_0=P_t$, $P_2=P_{\psi_2}$ and
$P_3=P_{\psi_5}$ are conserved. From (\ref{gml}), one obtains the following
explicit expressions for them
\ba\nl P_0=-\frac{(2l_pR)^2}{2\la^0}\cosh^2\rho,
\h P_2=-\sqrt{2}a\la^1\frac{(2l_pR)^2}{4\la^0}\sinh^2\rho,
\h P_3=-\sqrt{2}m\la^2\frac{(2l_pR)^2}{4\la^0}\sinh^2\rho.\ea

For compatibility of the membrane embedding with the constraints (\ref{cil}),
the conditions (\ref{cc}) must be fulfilled. In the present case, they lead to
$P_2=P_3=0$. This means that we have to work in the worldvolume gauge
$\la^i=0$. Then
\ba\nl P_2=P_3\equiv 0,\h \La^\mu_i P_\mu \equiv 0,\ea
and the constraints (\ref{cil}) are also identically satisfied.

In linear gauges, there is another consistency condition - (\ref{emr}), which
connect the membrane energy $E$ with all the conserved momenta $P_\mu$.
For the embedding, we are considering, (\ref{emr}) just states that
\ba\nl E=-VP_0=const.\ea
Thus, in the framework of the ansatz (\ref{a14}), the only nontrivial
conserved quantity is the M2-brane energy.

Finally, it remains to present the solution of the equations of
motion (\ref{eml}) and of the constraint (\ref{ecl}). Our
background (\ref{rb14}) depends on only one coordinate -
$x^1=\rho$. In this case, as explained in the appendix, the
constraint (\ref{ecl}) is first integral for the equation of
motion for $\rho(\tau)$, and the general solution satisfying
$\tau\left(\rho_0\right)=\tau_0$ is given by (\ref{tsol1}). For
the case at hand, it reads \ba\label{tsol14}
\tau\left(\rho\right)=\tau_0 + \int_{\rho_0}^{\rho}d\rho
\left[\frac{\la^0E}{(2\pi l_pR)^2}-\cosh^2\rho -
\left(2\la^0T_2\right)^2
(2l_pR)^2a^2m^2\sinh^4\rho\right]^{-1/2}.\ea

Let us now try to find a membrane solution based on more general embedding
of the type (\ref{tLA}), when the background seen by the M2-brane
depends on two coordinates. To this end, we choose the following ansatz
($X^\mu=X^{0,3,4}, X^a=X^{1,2}$ in our notations)
\ba\nl &&X^0(\tau,\delta,\s)\equiv t(\tau,\delta,\s)=\tau,
\\ \nl &&X^1(\tau,\delta,\s)=Y^1(\tau)=\rho(\tau),
\\ \label{la1} &&X^2(\tau,\delta,\s)=Y^2(\tau)=\psi_1(\tau),
\\ \nl &&X^3(\tau,\delta,\s)\equiv \psi_2(\tau,\delta,\s)=
\La_1^3\delta+\La_2^3\s,
\\ \nl &&X^4(\tau,\delta,\s)\equiv \psi_5(\tau,\delta,\s)=
\La_1^4\delta+\La_2^4\s .\ea
Comparing with (\ref{AdS7S4m}), one sees that the relevant background metric is
\ba\label{orb1} ds^2 = (2l_p R)^2\left[-\cosh^2\rho dt^2 + d\rho^2 +
\sinh^2\rho\left(d\psi_1^2 + \cos^2\psi_1 d\psi_2^2+\sin^2\psi_1
d\psi^2_5\right)\right].\ea

For the above membrane configuration, our action (\ref{olga}), in worldvolume
gauge $\la^i=0$, reads
\ba\label{oelg2} S^{LG}&=&\frac{(2\pi l_p R)^2}{\la^0}\int dt
\left[\dot{\rho}^2+\dot{\psi_1}^2\sinh^2\rho-\cosh^2\rho\right.
\\ \nl &-&\left.(2l_pR)^2\left(2\la^0T_2\Delta\right)^2
\sinh^4\rho\sin^2\psi_1\cos^2\psi_1\right],
\\ \nl &&\Delta=\La_1^3\La_2^4 - \La_1^4\La_2^3 .\ea
The corresponding Nambu-Goto type action is
\ba\label{nga2} S^{NG}=-(2\pi)^2(2l_pR)^3 T_2\Delta\int dt\sinh^2\rho
\sin\psi_1\cos\psi_1\sqrt{\cosh^2\rho-\dot{\rho}^2
- \dot{\psi_1}^2\sinh^2\rho}.\ea

According to (\ref{gml}) and (\ref{emr}), the conserved quantities are given by
\ba\nl E=-(2\pi)^2 P_0=\frac{(4\pi l_pR)^2}{2\la^0}\cosh^2\rho,\h
P_3=P_4=0.\ea
The compatibility conditions (\ref{cc}), and therefore - the constraints
(\ref{cil}), are identically satisfied. So, our next task is to solve
the equations of motion (\ref{eml}) and the remaining constraint (\ref{ecl}).
As far as the background (\ref{orb1}) is diagonal one, and
depends on two coordinates, we can use the general expressions
(\ref{fir}) and (\ref{fia}) for the first integrals of the
equations (\ref{eed}), which also solve the
constraint (\ref{ecd}), if the conditions
(\ref{r1}) and (\ref{r2}) are satisfied. Let us check if this is the case.
The conditions (\ref{r1}) are fulfilled, because they take the form
\ba\nl \mathcal{A}_a^L\equiv 0,\h \frac{\p}{\p\psi_1}
\sinh^2\rho\equiv 0.\ea
Consequently, it remains to satisfy the conditions
(\ref{r2}). In the case at hand, they
require, the right hand sides of (\ref{fir}) and (\ref{fia}) to
depend only on $Y^1=\rho$ and $Y^2=\psi_1$ respectively. To see if
this is true, let us write down the first integrals (\ref{fir})
and (\ref{fia}) explicitly
\ba\label{fir1}
&&\left(2l_pR\right)^4 \dot{\rho}^2 =
-\frac{D_2(\rho)}{\sinh^2\rho}\equiv F(\rho)\ge 0, \\ \label{fip1}
&&\left(2l_p R \sinh\rho\right)^4 \dot{\psi_1}^2 = D_2(\rho) +
\left(2l_pR \sinh\rho\right)^2\mathcal{U}^{L}(\rho,\psi_1)
\\ \nl &&= D_2(\rho) + \left(2l_pR \sinh\rho\right)^2
\left\{\frac{\la^0 E}{\pi^2} - \left(2l_pR\right)^2
\left[\cosh^2\rho \right.\right. \\ \nl &&+\left.\left.
(2l_pR)^2\left(2\la^0T_2\Delta\right)^2
\sinh^4\rho\sin^2\psi_1\cos^2\psi_1\right]\right\}.\ea It is
evident that the r.h.s. of the equation for $\dot{\rho}$ is a
function only on $\rho$, while the r.h.s. of the equation for
$\dot{\psi_1}$ is not a function only on $\psi_1$. Hence, the
second of the conditions (\ref{r2}) remains unsatisfied in the
general case. There exists, however, a particular case, when it
can be fulfilled. As long as the four parameters $\La^{3,4}_i$ in
our ansatz (\ref{la1}) are still arbitrary, we can restrict them
by the condition $\Delta=0$, and choose the arbitrary function
$D_2(\rho)$ as \ba\nl D_2(\rho)= d^2 - \left(2l_pR
\sinh\rho\right)^2 \left[\frac{\la^0 E}{\pi^2} -
\left(2l_pR\right)^2\cosh^2\rho\right]\le 0, \h d^2=const.\ea In
this way, the r.h.s. of (\ref{fip1}) became a constant and all
integrability conditions (\ref{r1}), (\ref{r2}), are satisfied.
\footnote{How the equations (\ref{fir1}) and (\ref{fip1}) can be
solved, we will explain on the example of the next case of
membrane embedding, considered below.}

The same result may be achieved by setting the membrane tension $T_2=0$,
instead of $\Delta=0$. In both cases, the solution of the equations
(\ref{fir1}) and (\ref{fip1}) will correspond to a {\it null membrane},
because the determinant of the worldvolume metric is zero for this
configuration. We note that such solution cannot be obtained by using the
Nambu-Goto type action (\ref{nga2}). It is identically zero in this case,
while the action (\ref{oelg2}), which we are using, simplifies to
\ba\nl S^{LG}_{0}&=&\frac{(2\pi l_p R)^2}{\la^0}\int dt
\left(\dot{\rho}^2+\dot{\psi_1}^2\sinh^2\rho-\cosh^2\rho\right).\ea

Let us turn to the more interesting case, when the M2-brane extends also
on the $S^4$-part of the $AdS_7\times S^4$ background. To this aim, we choose
the following embedding of type (\ref{tLA})
\footnote{The M-theory background 3-form on $S^4$ is zero for this ansatz.}
\ba\nl &&X^0(\tau,\delta,\s)\equiv t(\tau,\delta,\s)=
\La_0^0\tau+\La_1^0\delta+\La_2^0\s,
\\ \nl &&X^1(\tau,\delta,\s)=Y^1(\tau)=\rho(\tau),
\\ \label{la3} &&X^2(\tau,\delta,\s)\equiv \psi_2(\tau,\delta,\s)=
\La_0^2\tau+\La_1^2\delta+\La_2^2\s,
\\ \nl &&X^3(\tau,\delta,\s)\equiv \psi_5(\tau,\delta,\s)=
\La_0^3\tau+\La_1^3\delta+\La_2^3\s,
\\ \nl &&X^4(\tau,\delta,\s)=Y^4(\tau)=\alpha(\tau),
\\ \nl &&X^5(\tau,\delta,\s)\equiv \theta(\tau,\delta,\s)=
\La_0^5\tau+\La_1^5\delta+\La_2^5\s .\ea
The background seen by the membrane is ($\psi_1=\pi/4$)
\ba\label{orb2} ds^2 = (2l_p R)^2\left[-\cosh^2\rho dt^2 + d\rho^2 +
\frac{1}{2}\sinh^2\rho\left(d\psi_2^2+d\psi^2_5\right)
+\frac{1}{4}\left(d\alpha^2 + \cos^2\alpha d\theta^2\right)\right],\ea
and in our notations $X^\mu=X^{0,2,3,5}, X^a=X^{1,4}$.

For the ansatz (\ref{la3}) and in accordance with (\ref{emr}), the energy
$E$ is a linear combination of all conserved momenta $P_\mu$
\ba\nl E=-(2\pi)^2\La^\mu_0 P_\mu=
-\frac{(2\pi)^2}{2\la^0}\La_0^\mu\left(\La_0^\nu - \la^j\La_j^\nu\right)
g_{\mu\nu}.\ea
Actually, the compatibility conditions (\ref{cc}), (\ref{emr}), can be
satisfied by expressing three of the free parameters through the others.
If we choose to exclude $\La_1^5$, $\La_2^5$ and $P_0$, the following
equalities will hold
\ba\nl &&\La_1^5 = \frac{1}{\La_0^0 P_5}\left[\La_1^0
\left(E+\La_0^5 P_5\right)-D_{02}P_2-D_{03}P_3\right],
\\ \nl &&\La_2^5 = \frac{1}{\La_0^0 P_5}\left[\La_2^0
\left(E+\La_0^5 P_5\right)-d_{02}P_2-d_{03}P_3\right],
\\ \nl &&P_0=-\frac{1}{\La_0^0}\left(E
+\La_0^2P_2+\La_0^3P_3+\La_0^5P_5\right).\ea
Here
\ba\nl &&D_{02}=\La_0^0\La_1^2-\La_0^2\La_1^0,\h
D_{03}=\La_0^0\La_1^3-\La_0^3\La_1^0,
\\ \nl &&d_{02}=\La_0^0\La_2^2-\La_0^2\La_2^0,\h
d_{03}=\La_0^0\La_2^3-\La_0^3\La_2^0 .\ea

Our next task is to solve the equations of motion (\ref{eed}) and the
constraint (\ref{ecd}), where $\mathcal{A}_a=0$,
$\mathcal{G}_{aa}=(g_{11},g_{44})$,
\ba\nl \mathcal{U} &=& \frac{2\la^0 E}{(2\pi)^2} - \left(2\la^0T_2\right)^2
\det{(\La_i^\mu\La_j^\nu g_{\mu\nu})}
\\ \nl &=&\frac{2\la^0 E}{(2\pi)^2} + (2l_p R)^4\left(\la^0T_2\right)^2
\left\{\left[2\left(\Delta^2_{02}+\Delta^2_{03}\right)\cosh^2\rho
- \Delta^2_{23}\sinh^2\rho\right]\sinh^2\rho\right.
\\ \nl &+&\left.\frac{1}{2}\left[2\Delta^2_{05}\cosh^2\rho
- \left(\Delta^2_{25}+\Delta^2_{35}\right)\sinh^2\rho
\right]\cos^2\alpha\right\},\ea
and $\Delta_{\mu\nu}^2=\left(\La_1^\mu\La_2^\nu-\La_1^\nu\La_2^\mu\right)^2$.
Now, contrary to the previously considered case, we have enough freedom to
satisfy the integrability conditions (\ref{r1}) and (\ref{r2}), for arbitrary
value of the membrane tension. To this end, we can choose
\ba\nl &&\Delta^2_{25}+\Delta^2_{35}=2\Delta^2_{05},
\\ \nl &&D_4(\rho)=d - (l_pR)^2\left\{\frac{2\la^0 E}{(2\pi)^2}
+ (l_p R)^4\left(4\la^0T_2\right)^2\right.
\\ \nl &&\times \left.\left[2\left(\Delta^2_{02}
+\Delta^2_{03}\right)\cosh^2\rho
- \Delta^2_{23}\sinh^2\rho\right]\sinh^2\rho\right\}\le 0,\ea
where $d$ is arbitrary constant. After this choice, the first integrals
(\ref{fir}) and (\ref{fia}) of the
equations of motion for $\rho(\tau)$ and $\alpha(\tau)$ take the form
\ba\label{fir2}
&&\left(g_{11}\dot{\rho}\right)^2=-4D_4(\rho)\equiv \Phi_1(\rho)\ge 0,
\\ \label{fia2}
&&\left(g_{44}\dot{\alpha}\right)^2=
d+(l_pR)^6\left(4\la^0 T_2\Delta_{05}\right)^2\cos^2\alpha
\equiv \Phi_4(\alpha)\ge 0.\ea
The general solutions of these equations are given by
\ba\nl \tau(\rho)=(2l_pR)^2\int\frac{d\rho}{\sqrt{\Phi_1(\rho)}},\h
\tau(\alpha)=(l_pR)^2\int\frac{d\alpha}{\sqrt{\Phi_4(\alpha)}}.\ea
From (\ref{fir2}) and (\ref{fia2}), we can also find the orbit
$\rho=\rho(\alpha)$:
\ba\nl 4\int\frac{d\rho}{\sqrt{\Phi_1(\rho)}}=
\int\frac{d\alpha}{\sqrt{\Phi_4(\alpha)}}.\ea

\subsection{Exact membrane solutions in general gauges}

In this subsection, we will consider several M2-brane configurations in the
framework of the ansatz (\ref{tGA}), which corresponds to more general
embedding than (\ref{tLA}). Now, the membrane coordinates
$X^\mu(\tau,\delta,\s)$ are allowed to vary non-linearly with the proper time
$\tau$.

To begin with, let us take the most general ansatz of type (\ref{tGA}) for the
background (\ref{rb14})
\ba\nl &&X^0(\tau,\delta,\s)\equiv t(\tau,\delta,\s)=
\La_1^0\delta+\La_2^0\s+Y^0(\tau),
\\ \nl &&X^1(\tau,\delta,\s)=Y^1(\tau)=\rho(\tau),
\\ \label{ga1} &&X^2(\tau,\delta,\s)\equiv \psi_2(\tau,\delta,\s)=
\La_1^2\delta+\La_2^2\s+Y^2(\tau),
\\ \nl &&X^3(\tau,\delta,\s)\equiv \psi_5(\tau,\delta,\s)=
\La_1^3\delta+\La_2^3\s+Y^3(\tau),
\\ \nl &&X^\mu=X^{0,2,3},\h X^a=X^1.\ea
The conserved momenta are given in (\ref{gmgc}), and for our case they read
\ba\label{cmu} P_\mu = \frac{g_{\mu\nu}}{2\la^0}\left(\dot{Y}^\nu
- \la^j\La_j^\nu\right).\ea
In particular, the membrane energy is
\ba\label{ce} E=-p_0=-VP_0 = \frac{(4\pi l_pR)^2}{2\la^0}\cosh^2\rho
\left(\dot{Y}^0 - \la^j\La_j^0\right).\ea
The compatibility conditions (\ref{ccs}) are satisfied for
\ba\nl \La_1^3=\frac{1}{p_3}\left(\La_1^0 E - \La_1^2 p_2\right),\h
\La_2^3=\frac{1}{p_3}\left(\La_2^0 E - \La_2^2 p_2\right),\ea
and the following relations between the conserved quantities can be also
derived from them
\ba\nl E=\frac{\Delta_{23}}{\Delta_{03}}p_2
=-\frac{\Delta_{23}}{\Delta_{02}}p_3 .\ea

The background (\ref{rb14}) depends only on the $\rho$-coordinate. In this
case, the general solution for the membrane coordinate $\rho(\tau)$ is given
by (\ref{tsol1}), which for the case under consideration reduces to
\ba\nl \tau(\rho)=\tau_0 + \frac{1}{2\la^0}\int_{\rho_0}^{\rho}
\frac{d\rho}{\sqrt{W(\rho)}},\ea
where
\ba\nl W(\rho) &=& \frac{1}{(4\pi l_pR)^4}\left[\frac{E^2}{\cosh^2\rho}
-\frac{2(p_2^2+p_3^2)}{\sinh^2\rho}\right]
\\ \nl &-& (l_pR)^2\left(\frac{T_2\Delta_{02}}{p_3}\right)^2
\left[E^2\sinh^2\rho - 2(p_2^2+p_3^2)\cosh^2\rho\right]\sinh^2\rho.\ea
To see the difference between the membrane solutions, obtained in the
framework of different type of embeddings, one can compare the above result
with (\ref{tsol14}). Both solutions are for the same background (\ref{rb14}).

Working with the ansatz (\ref{ga1}), we have to write down also the solutions
for the remaining M2-brane coordinates $X^\mu$, given in the generic case in
(\ref{xmus1}). These general solutions are as follows
\ba\nl &&X^0(\rho,\delta,\s)\equiv t(\rho,\delta,\s)=
\La_1^0\left[\la^1\tau(\rho)+\delta\right]
+\La_2^0\left[\la^2\tau(\rho)+\s\right]
\\ \nl &&\hspace{4.5cm}+\frac{E}{(4\pi l_pR)^2}\int_{\rho_0}^{\rho}
\frac{d\rho}{\cosh^2\rho\sqrt{W(\rho)}},
\\ \nl &&X^2(\rho,\delta,\s)\equiv \psi_2(\rho,\delta,\s)=
\La_1^2\left[\la^1\tau(\rho)+\delta\right]
+\La_2^2\left[\la^2\tau(\rho)+\s\right]
\\ \nl &&\hspace{4.5cm}+\frac{2p_2}{(4\pi l_pR)^2}\int_{\rho_0}^{\rho}
\frac{d\rho}{\sinh^2\rho\sqrt{W(\rho)}},
\\ \nl &&X^3(\rho,\delta,\s)\equiv \psi_5(\rho,\delta,\s)=
\frac{1}{p_3}\left\{\left(\La_1^0 E - \La_1^2 p_2\right)
\left[\la^1\tau(\rho)+\delta\right] + \left(\La_2^0 E - \La_2^2 p_2\right)
\left[\la^2\tau(\rho)+\s\right]\right\}
\\ \nl &&\hspace{4.5cm}+\frac{2p_3}{(4\pi l_pR)^2}\int_{\rho_0}^{\rho}
\frac{d\rho}{\sinh^2\rho\sqrt{W(\rho)}}.\ea

The next M2-brane configuration, we will consider, is based on the most
general ansatz of type (\ref{tGA}) for the background (\ref{orb2})
\ba\nl &&X^0(\tau,\delta,\s)\equiv t(\tau,\delta,\s)=
\La_1^0\delta+\La_2^0\s+Y^0(\tau),
\\ \nl &&X^1(\tau,\delta,\s)=Y^1(\tau)=\rho(\tau),
\\ \label{ga2} &&X^2(\tau,\delta,\s)\equiv \psi_2(\tau,\delta,\s)=
\La_1^2\delta+\La_2^2\s+Y^2(\tau),
\\ \nl &&X^3(\tau,\delta,\s)\equiv \psi_5(\tau,\delta,\s)=
\La_1^3\delta+\La_2^3\s+Y^3(\tau),
\\ \nl &&X^4(\tau,\delta,\s)=Y^4(\tau)=\alpha(\tau),
\\ \nl &&X^5(\tau,\delta,\s)\equiv \theta(\tau,\delta,\s)=
\La_1^5\delta+\La_2^5\s+Y^5(\tau) ,
\\ \nl &&X^\mu=X^{0,2,3,5}, X^a=X^{1,4}.\ea
The expressions for the conserved momenta, and in particular for the membrane
energy are the same as in (\ref{cmu}) and (\ref{ce}).
The compatibility conditions (\ref{ccs}) are fulfilled identically, when
\ba\nl \La_1^5=\frac{1}{p_5}\left(\La_1^0 E-\La_1^2 p_2-\La_1^3 p_3\right),
\h \La_2^5=\frac{1}{p_5}\left(\La_2^0 E - \La_2^2 p_2-\La_2^3 p_3\right).\ea

As explained in appendix, we can now give three types of membrane
solutions: when $\alpha$ is fixed, when $\rho$ is fixed, and
without fixing any of the coordinates $\alpha$ and $\rho$, on
which the background (\ref{orb2}) depends. In the first two cases,
the formulas (\ref{tsol1}), (\ref{xmus1}) apply. In the last case,
we can use (\ref{fir}) and (\ref{fia}), if we succeed to satisfy
the integrability conditions (\ref{r1}), (\ref{r2}). In all these
cases, the effective scalar potential $\mathcal{U}^A(\rho,\alpha)$
is \footnote{The effective 1-form gauge potential
$\mathcal{A}^A=0$.} \ba\nl \mathcal{U}^A(\rho,\alpha)&=&
\frac{(\la^0)^2}{(2\pi)^4(l_pR)^2} \left[\frac{E^2}{\cosh^2\rho} -
\frac{2(p_2^2+p_3^2)}{\sinh^2\rho} -
\frac{4p_5^2}{\cos^2\alpha}\right]\\ \nl &+& (2l_p
R)^4\left(\la^0T_2\right)^2
\left\{\left[2\left(\Delta^2_{02}+\Delta^2_{03}\right)\cosh^2\rho
- \Delta^2_{23}\sinh^2\rho\right]\sinh^2\rho\right.
\\ \nl &+&\left.\frac{1}{2}\left[2\Delta^2_{05}\cosh^2\rho
- \left(\Delta^2_{25}+\Delta^2_{35}\right)\sinh^2\rho
\right]\cos^2\alpha\right\}.\ea

For $\alpha=\alpha_0=constant$, one obtains the solution
\ba\nl \tau\left(\rho\right)=\tau_0
+ 2l_pR\int_{\rho_0}^{\rho}
\frac{d\rho}{\sqrt{\mathcal{U}^A(\rho,\alpha_0)}},\ea
\ba \nl &&X^0(\rho,\delta,\s)\equiv t(\rho,\delta,\s)=
\La_1^0\left[\la^1\tau(\rho)+\delta\right]
+\La_2^0\left[\la^2\tau(\rho)+\s\right]
\\ \nl &&\hspace{4.5cm}+\frac{\la^0E}{(2\pi)^2l_pR}\int_{\rho_0}^{\rho}
\frac{d\rho}{\cosh^2\rho\sqrt{\mathcal{U}^A(\rho,\alpha_0)}},
\\ \nl &&X^2(\rho,\delta,\s)\equiv \psi_2(\rho,\delta,\s)=
\La_1^2\left[\la^1\tau(\rho)+\delta\right]
+\La_2^2\left[\la^2\tau(\rho)+\s\right]
\\ \nl &&\hspace{4.5cm}+\frac{2\la^0p_2}{(2\pi)^2 l_pR}\int_{\rho_0}^{\rho}
\frac{d\rho}{\sinh^2\rho\sqrt{\mathcal{U}^A(\rho,\alpha_0)}},
\\ \nl &&X^3(\rho,\delta,\s)\equiv \psi_5(\rho,\delta,\s)=
\La_1^3\left[\la^1\tau(\rho)+\delta\right]
+\La_2^3\left[\la^2\tau(\rho)+\s\right]
\\ \nl &&\hspace{4.5cm}+\frac{2\la^0p_3}{(2\pi)^2 l_pR}\int_{\rho_0}^{\rho}
\frac{d\rho}{\sinh^2\rho\sqrt{\mathcal{U}^A(\rho,\alpha_0)}},
\\ \nl &&X^5(\rho,\delta,\s)\equiv \theta(\rho,\delta,\s)=
\frac{1}{p_5}\left\{\left(\La_1^0 E-\La_1^2 p_2-\La_1^3 p_3\right)
\left[\la^1\tau(\rho)+\delta\right]\right.
\\ \nl &&\hspace{4.5cm}+\left.\left(\La_2^0 E - \La_2^2 p_2-\La_2^3 p_3\right)
\left[\la^2\tau(\rho)+\s\right]\right\}
\\ \nl &&\hspace{4.5cm}+\frac{2\la^0p_5}{(2\pi)^2 (l_pR)^2\cos^2\alpha_0}
\left[\tau(\rho)-\tau_0\right].\ea

For $\rho=\rho_0=constant$, the M2-brane solution is
\ba\nl \tau\left(\alpha\right)=\tau_0
+ l_pR\int_{\alpha_0}^{\alpha}
\frac{d\alpha}{\sqrt{\mathcal{U}^A(\rho_0,\alpha)}},\ea
\ba \nl &&X^0(\alpha,\delta,\s)\equiv t(\alpha,\delta,\s)=
\La_1^0\left[\la^1\tau(\alpha)+\delta\right]
+\La_2^0\left[\la^2\tau(\alpha)+\s\right]
+\frac{2\la^0E\left[\tau(\alpha)-\tau_0\right]}
{(4\pi l_pR)^2\cosh^2\rho_0}
\\ \nl &&X^2(\alpha,\delta,\s)\equiv\psi_2 (\alpha,\delta,\s)=
\La_1^2\left[\la^1\tau(\alpha)+\delta\right]
+\La_2^2\left[\la^2\tau(\alpha)+\s\right]
+\frac{4\la^0p_2\left[\tau(\alpha)-\tau_0\right]}
{(4\pi l_pR)^2\sinh^2\rho_0},
\\ \nl &&X^3(\alpha,\delta,\s)\equiv\psi_5 (\alpha,\delta,\s)=
\La_1^3\left[\la^1\tau(\alpha)+\delta\right]
+\La_2^3\left[\la^2\tau(\alpha)+\s\right]
+\frac{4\la^0p_3\left[\tau(\alpha)-\tau_0\right]}
{(4\pi l_pR)^2\sinh^2\rho_0},
\\ \nl &&X^5(\alpha,\delta,\s)\equiv \theta(\alpha,\delta,\s)=
\frac{1}{p_5}\left\{\left(\La_1^0 E-\La_1^2 p_2-\La_1^3 p_3\right)
\left[\la^1\tau(\alpha)+\delta\right]\right.
\\ \nl &&\hspace{4.5cm}+\left.\left(\La_2^0 E - \La_2^2 p_2-\La_2^3 p_3\right)
\left[\la^2\tau(\alpha)+\s\right]\right\}
\\ \nl &&\hspace{4.5cm}+\frac{2\la^0p_5}{(2\pi)^2 l_pR}\int_{\alpha_0}^{\alpha}
\frac{d\alpha}{\cos^2\alpha\sqrt{\mathcal{U}^A(\rho_0,\alpha)}}.\ea

When none of the coordinates $\rho$ and $\alpha$ is kept fixed, the conditions
(\ref{r1}), (\ref{r2}) will be fulfilled, if by using the arbitrariness of the
parameters $\La_i^\mu$ and of the function $D_4(\rho)$, we choose
\ba\nl &&\Delta^2_{25}+\Delta^2_{35}=2\Delta^2_{05},
\\ \nl &&D_4(\rho)=d - \frac{(\la^0)^2}{(2\pi)^4}\left[\frac{E^2}{\cosh^2\rho}
- \frac{2(p_2^2+p_3^2)}{\sinh^2\rho}\right]
-16(l_p R)^6\left(\la^0T_2\right)^2
\\ \nl &&\times \left[2\left(\Delta^2_{02}
+\Delta^2_{03}\right)\cosh^2\rho
- \Delta^2_{23}\sinh^2\rho\right]\sinh^2\rho\le 0,\ea
where $d$ is arbitrary constant. After this choice is made, the first integrals
(\ref{fir}) and (\ref{fia}) of the
equations of motion for $\rho(\tau)$ and $\alpha(\tau)$ reduce to
\ba\label{fir3}
&&\left(g_{11}\dot{\rho}\right)^2=
\frac{(2\la^0)^2}{(2\pi)^4}\left[\frac{E^2}{\cosh^2\rho}
+\frac{2(p_2^2+p_3^2)}{\sinh^2\rho}\right]
+(2l_p R)^6\left(\la^0T_2\right)^2
\\ \nl &&\times \left[2\left(\Delta^2_{02}
+\Delta^2_{03}\right)\cosh^2\rho
- \Delta^2_{23}\sinh^2\rho\right]\sinh^2\rho-4d \equiv F_1(\rho)\ge 0,
\\ \label{fia3}
&&\left(g_{44}\dot{\alpha}\right)^2=
d+(l_pR)^6\left(4\la^0 T_2\Delta_{05}\right)^2\cos^2\alpha
-\frac{(2\la^0p_5)^2}{(2\pi)^4\cos^2\alpha}
\equiv F_4(\alpha)\ge 0.\ea
The general solutions of the above two equations are
\ba\nl \tau(\rho)=(2l_pR)^2\int\frac{d\rho}{\sqrt{F_1(\rho)}},\h
\tau(\alpha)=(l_pR)^2\int\frac{d\alpha}{\sqrt{F_4(\alpha)}}.\ea
From (\ref{fir3}) and (\ref{fia3}), one can also find the orbit
$\rho=\rho(\alpha)$:
\ba\label{orb} 4\int\frac{d\rho}{\sqrt{F_1(\rho)}}=
\int\frac{d\alpha}{\sqrt{F_4(\alpha)}}.\ea

Now, we have to find the solutions for the remaining membrane coordinates
$X^\mu$. To this end, we will use (\ref{ymu}), i.e. the conservation laws for
$p_\mu$, which in our case read
\ba\nl \dot{Y}^\mu = \frac{2\la^0p_\nu}{(2\pi)^2}\left(g^{-1}\right)^{\mu\nu}
(\rho,\alpha)+\la^i\La_i^\mu.\ea
Representing $\dot{Y}^\mu$ as
\ba\nl \dot{Y}^\mu = \frac{\p Y^\mu}{\p\rho}\dot{\rho}
+ \frac{\p Y^\mu}{\p\alpha}\dot{\alpha},\ea
and using (\ref{fir3}) and (\ref{fia3}), one obtains
\ba\nl \frac{\sqrt{F_1(\rho)}}{(2l_pR)^2}\frac{\p Y^\mu}{\p\rho}
+ \frac{\sqrt{F_4(\alpha)}}{(l_pR)^2}\frac{\p Y^\mu}{\p\alpha}
= \frac{2\la^0p_\nu}{(2\pi)^2}\left(g^{-1}\right)^{\mu\nu}
(\rho,\alpha)+\la^i\La_i^\mu.\ea
This is a system of linear PDEs of first order, which general solution can
be easily found. Its replacement in the ansatz (\ref{ga2}), leads to the
following explicit expressions for the M2-brane coordinates $X^\mu$
\ba\nl &&X^0(\rho,\alpha;\delta,\s)\equiv t(\rho,\alpha;\delta,\s)=
\La_1^0\left[\la^1\tau(\rho)+\delta\right]
+\La_2^0\left[\la^2\tau(\rho)+\s\right]
\\ \nl &&\hspace{4.5cm}+\frac{2\la^0E}{(2\pi)^2}\int
\frac{d\rho}{\cosh^2\rho\sqrt{F_1(\rho)}} + f^0[C(\rho,\alpha)],
\\ \nl &&X^2(\rho,\alpha;\delta,\s)\equiv \psi_2(\rho,\alpha;\delta,\s)=
\La_1^2\left[\la^1\tau(\rho)+\delta\right]
+\La_2^2\left[\la^2\tau(\rho)+\s\right]
\\ \nl &&\hspace{4.5cm}+\frac{4\la^0p_2}{(2\pi)^2}\int
\frac{d\rho}{\sinh^2\rho\sqrt{F_1(\rho)}} + f^2[C(\rho,\alpha)],
\\ \nl &&X^3(\rho,\alpha;\delta,\s)\equiv \psi_5(\rho,\alpha;\delta,\s)=
\La_1^3\left[\la^1\tau(\rho)+\delta\right]
+\La_2^3\left[\la^2\tau(\rho)+\s\right]
\\ \nl &&\hspace{4.5cm}+\frac{4\la^0p_3}{(2\pi)^2}\int
\frac{d\rho}{\sinh^2\rho\sqrt{F_1(\rho)}} + f^3[C(\rho,\alpha)],
\\ \nl &&X^5(\rho,\alpha;\delta,\s)\equiv \theta(\rho,\alpha;\delta,\s)=
\frac{1}{p_5}\left\{\left(\La_1^0 E-\La_1^2 p_2-\La_1^3 p_3\right)
\left[\la^1\tau(\alpha)+\delta\right]\right.
\\ \nl &&\hspace{4.5cm}+\left.\left(\La_2^0 E - \La_2^2 p_2-\La_2^3 p_3\right)
\left[\la^2\tau(\alpha)+\s\right]\right\}
\\ \nl &&\hspace{4.5cm}+\frac{2\la^0p_5}{(2\pi)^2}\int
\frac{d\alpha}{\cos^2\alpha\sqrt{F_4(\alpha)}} + f^5[C(\rho,\alpha)],\ea
where $f^\mu[C(\rho,\alpha)]$ are arbitrary functions of $C(\rho,\alpha)$.
In turn, $C(\rho,\alpha)$ is the first integral of the equation (\ref{orb}).


\vspace*{.5cm} {\bf Acknowledgments} \vspace*{.2cm}\\ The author
would like to acknowledge the hospitality of the ICTP-Trieste,
where this investigation has been done. This work is supported by
the Abdus Salam International Center for Theoretical Physics,
Trieste, Italy, and by a Shoumen University grant under contract
{\it No.001/2003}.

\setcounter{section}{1}
\setcounter{subsection}{0}
\appendix{Generic formulas}

Here, we describe the membrane dynamics and find the corresponding solutions
of the equations of motion and constraints, in the framework of the two
ansatzes - (\ref{tLA}) and (\ref{tGA}).\footnote{We use part of the results
obtained in \cite{NPB656}, for the particular case $p=2$ (2-brane).}
Initially, the background fields $g_{MN}(x)$ and $b_{MNP}(x)$ are restricted
only by the conditions (\ref{ob}).

\subsection{Membranes dynamics in linear gauges}

In linear gauges, and under the conditions (\ref{ob}), the
action (\ref{oma}) reduces to (the over-dot is used for $d/d\tau$)
\ba\label{olga} &&S^{LG}=\int d\tau L^{LG}(\tau),\h V = \int d^2\xi
=\int d\delta d\s,\\
\nl &&L^{LG}(\tau)=
\frac{V}{4\lambda^0}\Bigl\{g_{ab}\dot{Y^a}\dot{Y^b} +
2\Bigl[\left(\Lambda_0^\mu-\lambda^i\Lambda_i^\mu\right)g_{\mu a}+
2\lambda^0T_2 B_{a12}\Bigr]\dot{Y^a}\\
\nl &&+\left(\Lambda_0^\mu-\lambda^i\Lambda_i^\mu\right)
\left(\Lambda_0^\nu-\lambda^j\Lambda_j^\nu\right)g_{\mu\nu}- \left(2\lambda^0
T_2\right)^2 \det(\Lambda_i^\mu\Lambda_j^\nu g_{\mu\nu})\\
\nl &&+ 4\lambda^0T_2\Lambda_0^{\mu}B_{\mu 12}\Bigr\},
\h B_{M12}\equiv b_{M\mu\nu}
\Lambda_1^{\mu}\Lambda_2^{\nu}.\ea

The constraints derived from the lagrangian (\ref{olga}) are:
\ba\label{c0l} &&g_{ab}\dot{Y^a}\dot{Y^b} +
2\left(\Lambda_0^\mu-\lambda^i\Lambda_i^\mu\right)g_{\mu
a}\dot{Y^a}+ \left(\Lambda_0^\mu-\lambda^i\Lambda_i^\mu\right) \\
\nl &&\times
\left(\Lambda_0^\nu-\lambda^j\Lambda_j^\nu\right)g_{\mu\nu}+\left(2\lambda^0
T_2\right)^2 \det(\Lambda_i^\mu\Lambda_j^\nu g_{\mu\nu})=0,\\
\label{cil} &&\Lambda_i^\mu\left[g_{\mu a}\dot{Y^a}+
\left(\Lambda_0^\nu-\lambda^j\Lambda_j^\nu\right)g_{\mu\nu}\right]=0.\ea

The lagrangian $L^{LG}$ does not depend on $\tau$ explicitly, so
the energy $E$ is conserved: \ba\nl
&&g_{ab}\dot{Y^a}\dot{Y^b}-\left(\Lambda_0^\mu-
\lambda^i\Lambda_i^\mu\right)\left(\Lambda_0^\nu-\lambda^j\Lambda_j^
\nu\right)g_{\mu\nu}+ \left(2\lambda^0 T_2\right)^2
\det(\Lambda_i^\mu\Lambda_j^\nu g_{\mu\nu})\\ \nl
&&-4\lambda^0T_2\Lambda_0^{\mu}B_{\mu 12}= \frac{4\lambda^0
E}{V}=constant.\ea With the help of the constraints (\ref{c0l})
and (\ref{cil}), one can replace this equality by the following
one \ba\label{p0l} \Lambda_0^\mu\left[g_{\mu a}\dot{Y^a}+
\left(\Lambda_0^\nu-\lambda^j\Lambda_j^\nu\right)g_{\mu\nu}
+2\lambda^0T_2 B_{\mu 12}\right]=-\frac{2\lambda^0 E}{V}.\ea

In linear gauges, the momenta $P_M$ take the form
\ba\label{gml} 2\la^0 P_M =g_{Ma}\dot{Y}^a+
\left(\Lambda_0^\nu-\lambda^j\Lambda_j^\nu\right)g_{M\nu}
+2\la^0T_2 B_{M12}.\ea
The comparison of (\ref{gml}) with (\ref{cil}) and (\ref{p0l}) gives
\ba\label{cc} \La^\mu_i P_\mu = constants = 0,
\\ \label{emr} \La^\mu_0 P_\mu=-\frac{E}{V} = constant.\ea
Therefore, in the linear gauges, the projections of the momenta
$P_\mu$ onto $\La^\mu_n$ are conserved. Moreover, as far as the
lagrangian (\ref{olga}) does not depend on the coordinates
$X^\mu$, the corresponding conjugated momenta $P_\mu$ are also
conserved.

The equalities (\ref{cc}) may be interpreted as solutions of the
constraints (\ref{cil}), which restrict the number of the
independent parameters in the theory.

Inserting (\ref{p0l}) and (\ref{cil}) into (\ref{c0l}), we obtain the
{\it effective} constraint
\ba\label{ecl} g_{ab}\dot{Y^a}\dot{Y^b}=\mathcal{U}^{L},\ea
where the {\it effective} scalar potential is given by
\ba\nl &&\mathcal{U}^{L}=-\left(2\lambda^0 T_2\right)^2
\det(\Lambda_i^\mu\Lambda_j^\nu g_{\mu\nu})+
\left(\Lambda_0^\mu-\lambda^i\Lambda_i^\mu\right)
\left(\Lambda_0^\nu-\lambda^j\Lambda_j^\nu\right)g_{\mu\nu}\\ \nl
&&+ 4\lambda^0\left(T_2 \Lambda_0^{\mu}B_{\mu 12}
+\frac{E}{V}\right).\ea

In the gauge $\lambda^m = constants$, the equations
of motion following from $L^{LG}$ take the form:
\ba\label{eml} g_{ab}\ddot{Y^b}
+\Gamma_{a,bc}\dot{Y^b}\dot{Y^c}=\frac{1}{2}\p_a \mathcal{U}^{L} +
2\p_{[a}\mathcal{A}_{b]}^{L}\dot{Y^b},\ea
where
\ba\nl \mathcal{A}_{a}^{L}=
\left(\Lambda_0^\mu-\lambda^i\Lambda_i^\mu\right)g_{a\mu}+
2\lambda^0T_2 B_{a12},\ea
is the {\it effective} 1-form gauge potential, generated by the non-diagonal
components $g_{a\mu}$ of the background metric and by the components
$b_{a\mu\nu}$ of the background 3-form gauge field.

\subsection{Membranes dynamics in general gauges}

We will use a superscript $A$ to denote that the corresponding quantity is
taken on the ansatz (\ref{tGA}). It is understood that the conditions
(\ref{ob}) are also fulfilled.

Now, the reduced lagrangian obtained from the action (\ref{oma})
is given by \ba\nl &&L^{A}(\tau)=
\frac{V}{4\lambda^0}\left[g_{MN}\dot{Y^M}\dot{Y^N} -
2\left(\lambda^i\Lambda_i^\mu g_{\mu N}- 2\lambda^0T_2
B_{N12}\right)\dot{Y^N}\right.\\ \nl
&&+\left.\lambda^i\Lambda_i^\mu\lambda^j\Lambda_j^\nu g_{\mu\nu}-
\left(2\lambda^0 T_2\right)^2 \det(\Lambda_i^\mu\Lambda_j^\nu
g_{\mu\nu})\right].\ea The constraints, derived from the above
lagrangian, are: \ba\label{c0g} &&g_{MN}\dot{Y^M}\dot{Y^N} -
2\lambda^i\Lambda_i^\mu g_{\mu N}\dot{Y^N}+
\lambda^i\Lambda_i^\mu\lambda^j\Lambda_j^\nu
g_{\mu\nu}+\left(2\lambda^0 T_2\right)^2
\det(\Lambda_i^\mu\Lambda_j^\nu g_{\mu\nu})=0,\\ \label{cig}
&&\Lambda_i^\mu\left(g_{\mu N}\dot{Y^N} -\lambda^j\Lambda_j^\nu
g_{\mu\nu}\right)=0.\ea The corresponding momenta are ($P_M =
p_M/V$) \ba\nl 2\la^0 P_M =g_{MN}\dot{Y}^N-\lambda^j\Lambda_j^\nu
g_{M\nu} +2\la^0T_2 B_{M12},\ea and part of them, $P_\mu$, are
conserved \ba\label{gmgc} g_{\mu
N}\dot{Y}^N-\lambda^j\Lambda_j^\nu g_{\mu\nu} +2\la^0T_2 B_{\mu
12}= 2\la^0 P_\mu = constants ,\ea because $L^{A}$ does not depend
on $X^\mu$. From (\ref{cig}) and (\ref{gmgc}), the compatibility
conditions follow \ba\label{ccs} \La^\mu_i P_\mu =0.\ea We will
regard on (\ref{ccs}) as a solution of the constraints
(\ref{cig}), which restricts the number of the independent
parameters $\La^\mu_i$. That is why from now on, we will deal only
with the constraint (\ref{c0g}).

In the gauge $\lambda^m = constants$, the equations
of motion for $Y^N$, following from $L^{A}$, have the form
\ba\label{emg} g_{LN}\ddot{Y^N}
+\Gamma_{L,MN}\dot{Y^M}\dot{Y^N}=\frac{1}{2}\p_L \mathcal{U}^{in} +
2\p_{[L}\mathcal{A}_{N]}^{in}\dot{Y^N},\ea
where
\ba\nl &&\mathcal{U}^{in}=-\left(2\lambda^0 T_2\right)^2
\det(\Lambda_i^\mu\Lambda_j^\nu g_{\mu\nu})+
\lambda^i\Lambda_i^\mu\lambda^j\Lambda_j^\nu g_{\mu\nu},\\
\nl &&\mathcal{A}_{N}^{in}=
-\lambda^i\Lambda_i^\mu g_{N\mu}+
2\lambda^0T_2 B_{N12}.\ea
Let us first consider this part of the equations of motion (\ref{emg}), which
corresponds to $L=\la$. It is easy to check that they just express the fact
that the momenta $P_\mu$ are conserved. Therefore, we have to deal only
with the other part of the equations of motion, corresponding to $L=a$
\ba\label{emga} g_{aN}\ddot{Y^N}
+\Gamma_{a,MN}\dot{Y^M}\dot{Y^N}=\frac{1}{2}\p_a \mathcal{U}^{in} +
2\p_{[a}\mathcal{A}_{N]}^{in}\dot{Y^N}.\ea

Our next task is to eliminate the variables $\dot{Y^\mu}$ from these equations
and from the constraint (\ref{c0g}). To this
end, we will use the conservation laws (\ref{gmgc}) to express
$\dot{Y^\mu}$ through $\dot{Y^a}$. The result is \ba\label{ymu}
\dot{Y^\mu}=\left(g^{-1}\right)^{\mu\nu} \left[2\la^0(P_\nu - T_2 B_{\nu
12}) -g_{\nu a}\dot{Y^a}\right]+\la^i\La^\mu_i.\ea
By using (\ref{ymu}), after some calculations, one rewrites
the equations of motion (\ref{emga}) and the constraint (\ref{c0g}) in the
form \ba\nl &&h_{ab}\ddot{Y}^b +
\Gamma^{\bf{h}}_{a,bc}\dot{Y}^b\dot{Y}^c = \frac{1}{2}\p_a \mathcal{U}^{A}
+ 2\p_{[a}\mathcal{A}^{A}_{b]}\dot{Y}^b,\\ \nl
&&h_{ab}\dot{Y}^a\dot{Y}^b = \mathcal{U}^{A},\ea where a new, {\it
effective metric} appeared \ba\nl h_{ab} = g_{ab} -
g_{a\mu}(g^{-1})^{\mu\nu}g_{\nu b}.\ea $\Gamma^{\bf{h}}_{a,bc}$ is the
connection compatible with this metric \ba\nl
\Gamma^{\bf{h}}_{a,bc}=\frac{1}{2}\left(\p_b h_{ca} +\p_c h_{ba}-\p_a
h_{bc}\right).\ea The new, {\it effective} scalar and gauge potentials are
given by \ba\nl &&\mathcal{U}^{A}=-\left(2\lambda^0 T_2\right)^2
\det(\Lambda_i^\mu\Lambda_j^\nu g_{\mu\nu}) - (2\la^0)^2 \left(P_\mu-T_2
B_{\mu 12}\right)\left(g^{-1}\right)^{\mu\nu} \left(P_\nu-T_2
B_{\nu 12}\right),\\ \nl &&\mathcal{A}_{a}^{A}=
2\lambda^0\left[g_{a\mu}\left(g^{-1}\right)^{\mu\nu} \left(P_\nu-T_2
B_{\nu 12}\right)+ T_2 B_{a12}\right].\ea

\subsection{Solutions of the equations of motion}

The two cases of membrane dynamics considered so far, have one common feature.
The dynamics of the
corresponding reduced particle-like system is described by {\it effective}
equations of motion and one {\it effective} constraint, which have the
{\it same form}, independently of the ansatz used to reduce the membranes
dynamics. Our aim here is to give their exact solutions.
To be able to describe the two cases simultaneously, let us first introduce
some general notations.

We will search for solutions of the following system of {\it nonlinear}
differential equations \ba\label{ee} &&\mathcal{G}_{ab}\ddot{Y}^b +
\Gamma^{\mathcal{G}}_{a,bc}\dot{Y}^b\dot{Y}^c = \frac{1}{2}\p_a
\mathcal{U} + 2\p_{[a}\mathcal{A}_{b]}\dot{Y}^b,\\ \label{ec}
&&\mathcal{G}_{ab}\dot{Y}^a\dot{Y}^b = \mathcal{U},\ea where
$\mathcal{G}_{ab}$, $\Gamma^{\mathcal{G}}_{a,bc}$, $\mathcal{U}$, and
$\mathcal{A}_a$ can be as follows \ba\nl
\mathcal{G}_{ab}=\left(g_{ab},h_{ab}\right),\h
\Gamma^{\mathcal{G}}_{a,bc}=\left(\Gamma_{a,bc},\Gamma^{\bf{h}}_{a,bc}\right),
\h \mathcal{U}=\left(\mathcal{U}^{L},\mathcal{U}^{A}\right),
\h \mathcal{A}_a=\left(\mathcal{A}_{a}^{L},\mathcal{A}_{a}^{A}\right),\ea
depending on the membrane embedding.

Let us start with the simplest case, when the background fields depend on
only one coordinate $X^a=Y^a(\tau)$.
In this case the solution of (\ref{ee}),
compatible with (\ref{ec}), is just the constraint (\ref{ec}). In other
words, (\ref{ec}) is first integral of the equation of motion for the
coordinate $Y^a$. By integrating (\ref{ec}), one obtains the following
exact membrane solution \ba\label{tsol1}\tau\left(X^a\right)=\tau_0
+ \int_{X_0^a}^{X^a}
\left(\frac{\mathcal{U}}{\mathcal{G}_{aa}}\right)^{-1/2}dx,\ea where
$\tau_0$ and $X_0^a$ are arbitrary constants.

When one works in the framework of the general ansatz (\ref{tGA}),
one has to also write down the solution for the remaining coordinates $X^\mu$.
It can be obtained as follows. One represents $\dot{Y}^\mu$ as
\ba\nl \dot{Y}^\mu =\frac{dY^\mu}{dY^a}\dot{Y}^a,\ea
and use this and (\ref{ec}) in (\ref{ymu}). The result is a system of ordinary
differential equations of first order with separated variables, which
integration is straightforward. Replacing the obtained solution for
$Y^\mu(X^a)$ in the ansatz (\ref{tGA}), one finally arrives at
\ba\label{xmus1} &&X^\mu(X^a,\xi^i)=
\La^\mu_i\left[\la^i\tau(X^a)+\xi^i\right]
\\ \nl &&+\int_{X^a_0}^{X^a}\left(g^{-1}\right)^{\mu\nu}
\left[2\la^0(P_\nu-T_2 B_{\nu 12})
\left(\frac{\mathcal{U}^A}{h_{aa}}\right)^{-1/2} - g_{\nu a}\right]dx.\ea

Let us turn to the more complicated case, when the background fields depend on
more than one coordinate $X^a=Y^a(\tau)$. If the metric
$\mathcal{G}_{ab}$ is a diagonal one, then the effective
equations of motion (\ref{ee}) and the effective constraint (\ref{ec})
can be rewritten in the form
\ba\label{eed} &&\frac{d}{d\tau}\left(\mathcal{G}_{aa}\dot{Y}^a\right)^2 -
\dot{Y}^a\p_a\left(\mathcal{G}_{aa}\mathcal{U}\right)
\\ \nl &&+ \dot{Y}^a\sum_{b\ne a}
\left[\p_a\left(\frac{\mathcal{G}_{aa}}{\mathcal{G}_{bb}}\right)
\left(\mathcal{G}_{bb}\dot{Y}^b\right)^2
- 4\p_{[a}\mathcal{A}_{b]}\mathcal{G}_{aa}\dot{Y}^b\right] = 0,
\\ \label{ecd}
&&\mathcal{G}_{aa}\left(\dot{Y}^a\right)^2
+\sum_{b\ne a}\mathcal{G}_{bb}\left(\dot{Y}^b\right)^2 = \mathcal{U}.\ea

To find solutions of the above equations without choosing particular
background, we can fix all coordinates $X^a$ except one. Then the exact
membrane solution of the equations of motion is given
again by the same expression (\ref{tsol1}) for $\tau\left(X^a\right)$.
In the case when one is using the general ansatz (\ref{tGA}),
the solution (\ref{xmus1}) still also hold.

To find solutions depending on more than one coordinate, we have to impose
further restrictions on the background fields.
We cannot give a prescription how to solve the problem in the general case.
However, we can give an example of sufficient conditions, which are fulfilled
in many cases, and which allow us to find the first integrals of the equations
of motion (\ref{eed}), compatible with the effective constraint (\ref{ecd}).
If we denote one of the coordinates $Y^a$ with $Y^r$ and $Y^{\alpha}$
are the others, these conditions on the background can be written as
\ba\label{r1} \mathcal{A}_a\equiv
(\mathcal{A}_r,\mathcal{A}_\alpha)= (\mathcal{A}_r,\p_\alpha f),\h
\p_\alpha\left(\frac{\mathcal{G}_{\alpha\alpha}}
{\mathcal{G}_{aa}}\right)=0,\\ \label{r2}
\h\p_\alpha\left(\mathcal{G}_{rr}\dot{Y}^r\right)^2=0,\h
\p_r\left(\mathcal{G}_{\alpha\alpha}\dot{Y}^\alpha\right)^2=0.\ea
By using the restrictions given above, one obtains the following
first integrals of the equations (\ref{eed}), which also solve the
constraint (\ref{ecd})
\ba\label{fir} \left(\mathcal{G}_{rr}\dot{Y}^r\right)^2 =
\mathcal{G}_{rr} \left[(1-n_\alpha)\mathcal{U}
-2n_\alpha\left(\mathcal{A}_r-\p_r f\right)\dot{Y}^r
-\sum_{\alpha}\frac{D_{\alpha} \left(Y^{a\ne\alpha}\right)}
{\mathcal{G}_{\alpha\alpha}}\right]= F_r\left(Y^r\right)\ge 0,\\
\label{fia}
\left(\mathcal{G}_{\alpha\alpha}\dot{Y}^\alpha\right)^2 =D_{\alpha}
\left(Y^{a\ne\alpha}\right) + \mathcal{G}_{\alpha\alpha}\left[\mathcal{U}
+2\left(\mathcal{A}_r-\p_r f\right)\dot{Y}^r\right]= F_{\alpha}
\left(Y^{\beta}\right)\ge 0,\ea where $n_\alpha$ is the
number of the coordinates $Y^\alpha$, and $D_{\alpha}$, $F_r$, $F_{\alpha}$
are arbitrary functions of their arguments.

Further progress is possible, when working with particular background
configurations, allowing for separation
of the variables in (\ref{fir}) and (\ref{fia}).



\bigskip

\begin{thebibliography}{}

\bibitem{4} S. S. Gubser, I. R. Klebanov and A. M. Polyakov, {\it A
semi-classical limit of the gauge/string correspondence}, Nucl.
Phys. {\bf B 636} (2002) 99-114 [hep-th/0204051].
\bibitem{6} E. Sezgin and P. Sundell, {\it Massless higher spins and
holography}, Nucl. Phys. {\bf B 644} (2002) 303-370
[hep-th/0205131].
\bibitem{10} M. Alishahiha and M. Ghasemkhani, {\it Orbiting membranes in
M-theory on $AdS_7\times S^4$ background}, JHEP 08 (2002) 046
[hep-th/0206237].
\bibitem{14} M. Alishahiha and A. E. Mosaffa, {\it Circular semiclassical
string solutions on confining $AdS/CFT$ backgrounds}, JHEP 10
(2002) 060 [hep-th/0210122].
\bibitem{27} S. A. Hartnoll and C. Nunez, {\it Rotating membranes on $G_2$
manifolds, logarithmic anomalous dimensions and $\mathcal{N}=1$
duality}, JHEP 02 (2003) 049 [hep-th/0210218].
\bibitem{NPB656} P. Bozhilov, {\it Probe branes dynamics: exact solutions in
general backgrounds}, Nucl. Phys. {\bf B 656} (2003) 199-225
[hep-th/0211181].




\end{thebibliography}
\end{document}